\documentclass{jfm}
\usepackage[dvips]{graphicx}    

\begin{document} 

\title[Friction law for dense granular flows: application to the 
motion of a mass down a rough inclined plane.]
{Friction law for dense granular flows: application to the 
motion of a mass down a rough inclined plane.}

\author[O. Pouliquen and Y. Forterre] 
{O\ls L\ls I\ls V\ls I\ls E\ls R\ns  P\ls O\ls
U\ls L\ls I\ls Q\ls U\ls E\ls N\ns \and
Y\ls O\ls \"E\ls L\ns \ns F\ls O\ls R\ls T\ls E\ls R\ls R\ls E\ls }

\affiliation{IUSTI, UniversitŽ de Provence, CNRS UMR 6595,\\ 5 rue 
Enrico Fermi, 13453 Marseille cedex 13, France.} 

\date{2000} 

\maketitle

\begin{abstract}

The problem of the spreading of a granular mass released at the top of 
a rough inclined plane was investigated.  We experimentally measure 
the evolution of the avalanche from the initiation up to the deposit 
using a MoirŽ image processing technique.  The results are 
quantitatively compared with the prediction of an hydrodynamic model 
based on depth averaged equations.  In the model, the interaction 
between the flowing layer and the rough bottom is described by a non 
trivial friction force whose expression is derived from measurements 
on steady uniform flows.  We show that the spreading of the mass is 
quantitatively predicted by the model when the mass is released on a 
plane free of particles.  When an avalanche is triggered on an 
initially static layer, the model fails in quantitatively predicting 
the propagation but qualitatively captures the evolution.

\end{abstract}

\section{Introduction}

Gravity driven geophysical events sometimes involve the flow of a dry
granular material.  Landslides, rock avalanches and pyroclastic flows are 
examples of natural hazards where a granular mass flows down a slope.  
One of the main issues is to predict the flow trajectory over the 
complex topography, the velocity  and the runout 
distance in order to define a safety zone.  

The difficulty in describing granular geophysical flows lies in the 
uncertainty in the constitutive equations for the flow of dry granular 
media.  Constitutive equations are known for rapid granular flows in a 
low concentration regime, when the particles interactions are 
dominated by the collisions .  For this regime, a kinetic theory can 
be developed based on the kinetic theory of dense gases taking into 
account the dissipation during the collisions 
(\cite{haff83,lun84,azanza99,goldhirsch99}).  However, for dense flows 
when the particles not only undergo collisions but also friction and 
multicontact interactions with neighboring particles, the kinetic 
approach is no longer valid.  Some attempts have been made to 
incorporate into the kinetic theory which describes the collisional 
interactions, an empirical rate independent stress tensor in order to 
take into account the friction (\cite{savage83,johnson90,anderson92}).  
However, it is not clear that these approaches are valid for dense 
flows when multi particles contacts are present.  More recently, Savage (1998) 
proposed an analysis for high concentration flows based on a fluctuating 
plasticity model which has not yet been applied for flow down slopes.  
Another approach has been suggested by Mills {\it et al} (1999,2000) 
which is based on the idea that force chains are present in the media 
inducing a non local formulation of the stresses. 

A promising approach for describing unsteady and non uniform flow on 
complex geometry such as geophysical flows, is the depth averaged St 
Venant (1871) approach .  In this framework, the material is assumed 
to be incompressible and the mass and momentum equations are written 
in a depth averaged form.  This analysis is valid under the assumption 
that the flowing layer is thin compared to its lateral extension which 
is often the case for geophysical flows.  Depth averaging allows to 
avoid a complete 3D description of the flow: the complex rheology of 
the granular material is incorporated in a single term describing the frictional 
stress that develops at the interface between the flowing material and 
the rough surface.  Our goal in this paper is to propose an expression 
for this friction force based on experimental measurements, and to 
quantitatively compare the prediction of the depth averaged approach 
with well controlled experiments of granular avalanches.

Depth averaged equations have been introduced in the context of 
granular flows by Savage and Hutter (1989).  In their model, 
the interaction between the granular material and the rough surface is 
described by a simple friction law: the shear stress at the bottom is 
proportional to the normal stress, the coefficient of friction being 
a constant.  The model works well when the surface of the plane is 
smooth enough: Savage, Hutter and coworkers were able to predicts the motion and 
spreading of a granular mass on steep slopes in two and three 
dimensions 
(\cite{savage89,wieland99,gray99}) . Experiments have been also carried out on 
curved beds (\cite{greve93,greve94,koch94}), and the measurements agree relatively well with the 
prediction of the depth averaged model. However, all the experiments 
have been carried out with high inclinations (higher than $40^{o}$) 
and with relatively smooth beds (typically 3mm beads rolling on flat 
surfaces or sand paper). 

For the flow of granular material on rough surfaces (where the 
roughness of the bed is of the order of the particle size) 
and for moderate inclinations, the description in term of a simple 
solid friction no longer holds. This is shown by several 
experimental observations:

a) First, it has been shown that steady uniform flows are observed in a 
 range of inclination angles 
(\cite{suzuki71,hungr84,vallance94,ancey96,pouliquen99a}).  This implies that the bed shear 
stress is not a simple solid friction but has a velocity dependence 
in order to balance the gravity for different inclinations.

b) Secondly, the onset of flow on rough inclined planes is not 
precisely described by the simple solid friction law.  According to this law, 
the flow is possible as soon as the inclination is higher than the 
friction angle.  Experiments reveal a more complex behavior: the onset 
of flow depends both on the inclination of the plane and on the 
thickness of the granular layer 
(\cite{pouliquen96,daerr99,pouliquen99a}).  A thin layer has more 
difficulty to flow than a thick one.

c) Thirdly, an hysteresis is observed between flowing and static 
states. A 
static layer of material resting on a rough surface will start flowing 
at a given threshold of the inclination. However it will not stop before the 
slope decreases below a second threshold. In between these two angles 
there exist metastable states, where avalanches can be triggered by 
perturbations. The complex and rich dynamics of the avalanches has been 
recently studied by Daerr and Douady (1999) and Daerr (2001).

All these features can not be captured by the simple solid friction 
assumption.  The question we want to address in this study is the 
following: can we find a more realistic friction law within the depth 
averaged framework which can capture the behavior of thin granular 
layers on rough planes?  In a previous work (\cite{pouliquen99a}), we 
have proposed an empirical friction law based on scaling properties 
measured for steady uniform flows.  We have shown that a depth 
averaged description taking into account this friction predicts the 
steady shape of granular fronts propagating down a slope 
(\cite{pouliquen99b}).  However, unsteady flows where static material 
start to move, or moving material comes to rest, involve a complex 
dynamics which is not a priori taking into account in the friction 
law.  In the present study our goal is to check to what extent the 
empirical friction law found for steady uniform flows can be used in 
the framework of the depth averaged equations to {\it quantitatively} 
predict unsteady and non uniform flows where deposit are formed and 
avalanches are triggered.

To this end, we have investigated two different experimental 
configurations.  The first one is the spreading of a granular mass 
released at the top of a rough inclined plane.  The second is the 
release of a mass on a static layer of grains initially present on the 
rough plane.  With the first configuration we can study how the 
material comes to rest and creates a deposit, whereas with the second 
we can study how static material is put into motion by flowing 
material.  In both configurations the time evolution of the mass is 
measured and compared with the predictions of the depth averaged model.

The experimental setup and the measurement procedure are presented 
in Section 2. In Section 3 the depth averaged model is described and 
the choice of the friction law is discussed. The results are presented 
in Section 4. Discussion about the sensitivity of the model to the 
parameters is given in Section 5 before concluding about the relevance and limits 
of the depth averaged approach in Section 6. 

\section{Experimental configurations}

\subsection{Experimental setup}

The experimental setup is a 2m long and 70cm wide rough plane whose 
inclination $\theta$ can be precisely controlled.  The surface is made 
rough by gluing one layer of particles on the plane.  The particles we 
used are glass beads 0.5mm $\pm 0.04$ in diameter.  The first set of 
experiments consists in removing a spherical cap full of beads at the 
top of the plane.  The released material then starts flowing down the 
slope, the mass spreads and ultimately stops leaving a tear shaped 
deposit (Fig.  1a).  Three roughly homothetic spherical caps have been 
used: a small one ($r=2.1$ cm $R=3.7$ cm), a medium one ($r=3.1$ cm 
$R=6$ cm) and a large one ($r=4$ cm $R=8$ cm, see Fig.  1a for the 
definition of $r$ and $R$ ).  In order to precisely control the 
amount of material present in the cap, the mass of beads poured in the 
cap is initially weighed before each run (62 g for the small cap, 231 
g for the medium one, 524 g for the large cap).

\begin{figure}
\centering 
\includegraphics[scale=0.6]{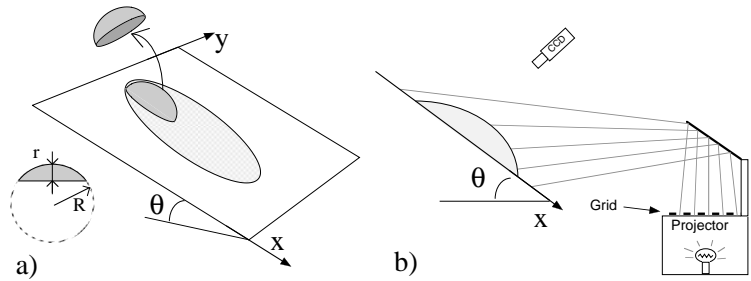}
\caption{a) Sketch of the experimental set-up. The spherical caps we 
used are 
part of a sphere of radius R. b) Visualisation method.}
\end{figure}

In a second set of experiments the mass is released on the top of a layer 
of material already lying on the surface.  The initially static 
layer is prepared by creating a steady uniform flow over the whole 
surface and by suddenly stopping the supply. In this case, a uniform 
deposit is created with a well defined thickness $h_{stop}$ depending 
only on the inclination $\theta$ (\cite{pouliquen99a}).  The spherical cap is then 
carefully placed on the layer and filled with the beads.  When 
the cap is suddenly removed, the mass spreads out putting into 
motion some of the initially static material, triggering an 
avalanching front.

\subsection{Measurement method}

In order to measure precisely the time evolution of the granular mass, 
we have developed a method inspired by a Moir\'e 
method (\cite{sansoni99}).  A grid pattern made of horizontal lines is 
projected on the plane by an overhead projector as sketched in Fig.  
1b.  The projection angle is small enough such that the presence of 
the granular mass on the surface induces a significant deformation of 
the grid (Fig.  2b).  Pictures of the plane are recorded by a CCD 
camera positioned at the vertical of the plane.  The local shift of 
the lines observed between the deformed pattern in the presence of a 
granular mass (Fig.2b) and the initial regular pattern when the 
surface is flat (Fig.2a) is proportional to the local thickness of the 
granular layer.  In order to quantitatively get the thickness $h(x,y)$ 
we proceed as follows.  The reference picture Fig.  2a and the picture 
to be analyzed Fig.  2b are digitalized which gives two real amplitudes 
$A_{ref}(x,y)$ and $A(x,y)$.  The 2D spatial Fourier transforms of 
both amplitudes are then computed.  Both pictures being close to a 
regular pattern, the Fourier transforms are found to present well 
defined peaks at the complex wavenumbers $(2\pi/\lambda,0)$ and 
$(-2\pi/\lambda,0)$, $\lambda$ being the wavelength of the projected 
grid.  For Fig.  2b, the information about the slight deviation from 
the reference pattern induced by the granular mass is contained in the 
width of the peak.  In order to extract the phase of the picture, the 
right half of the wavenumber spectrum ($k_{x}>0$) is used to 
reconstruct two complex amplitudes $A^{c}_{ref}(x,y)$ and 
$A^{c}(x,y)$.  The phase $\phi_{ref}(x,y)$ and $\phi(x,y)$ of 
$A^{c}_{ref}$ and $A^{c}$ gives the phase of the pattern in picture 
2a and 2b.  The thickness $h(x,y)$ is then simply proportional to the 
phase difference $\phi(x,y)-\phi_{ref}(x,y)$.  The coefficient of 
proportionality is found by measuring the 
phase shift induced by a 1cm thick plate. The Figure 2c shows the 
contours of constant thickness given by the analysis.

  \begin{figure}
  \centering 
  \includegraphics[scale=0.5]{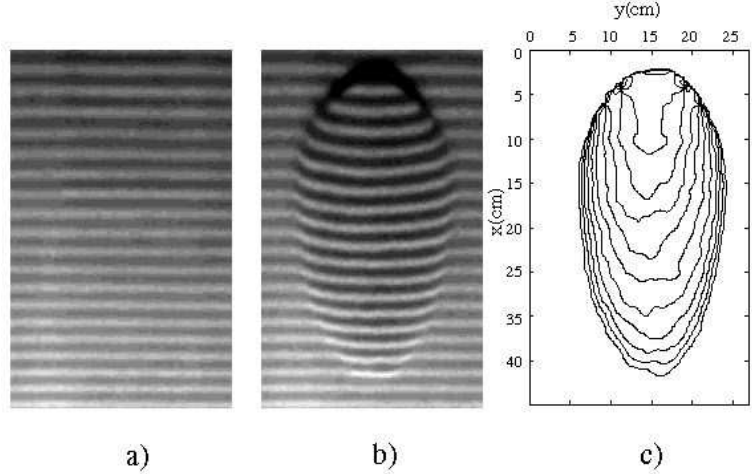}
  \caption{a) Reference picture; b) Picture to be analyzed; c) 
  Contours of  constant thickness every 0.5mm given by the phase difference 
  between picture a) and b).}
  \end{figure}

The time evolution of the free surface of the mass $h(x,y,t)$ is 
obtained by analyzing by this method each image of a movie recorded 
during the flow.  The thickness measurement is estimated to be precise 
up to $\pm 0.3$mm.  The phase measurement method breaks down when 
shadows are present behind the mass as in Fig 2b.  In the region close 
to the shadow, the thickness is not correctly estimated and the 
contour of the deposit is not correct.  The contour in this region is 
then estimated from direct visualisation of the pictures.  However, 
shadows are observed only at the first stage of the spreading when the 
layer is thick.

\section{Theoretical model}
 
\subsection{Depth averaged equations}

The use of depth averaged equations to describe the flows under 
consideration here is motivated by the large aspect ratio of 
the granular mass: its lateral extension is large compare to the thickness of the layer. 

More precisely, in order to derive the St Venant equations we assume 
that the variation of the flow 
takes place on lengths much larger than the thickness.  Assuming in 
addition that the flow is incompressible of constant density $\rho$, 
which is justified for the dense flow regim studied here 
\cite{savage89,ertas01}, 
we obtain the depth averaged mass and momentum conservations by 
integrating the full 3D equations (\cite{savage89}). The following 
equations are written in terms of the local thickness $h (x, y, t)$ 
and the depth averaged velocity ${\bf u} (x, y, t)=u {\bf e}_x + v{\bf 
e}_y$:

\begin{eqnarray}
\displaystyle\frac{\partial  h}{\partial t} + 
\displaystyle\frac{\partial  (hu)}{\partial x}+
\displaystyle\frac{\partial  (hv)}{\partial y} &=& 0 ,
\label{masscons}\\
 \rho \left( \displaystyle\frac{\partial (h {\bf u})}{\partial 
t} + \alpha {\bf 
\nabla} \cdot \left( h {\bf u} \otimes {\bf u} \right) \right) &=&  
\rho g h \cos \theta \left( \tan \theta {\bf e}_x - \mu \left( h, \|{\bf u}\| \right) 
\displaystyle\frac{{\bf u}} { \|{\bf u}\|} - K {\bf 
\nabla} h \right) ,
\label{mom}
\end{eqnarray}

Equation \ref{masscons} is the mass conservation.  Equation \ref{mom} 
is the momentum conservation where the acceleration is balanced by 
three forces.  In the acceleration expression, the coefficient 
$\alpha$ is related to the assumed velocity profile across the layer.  
In the case of a plug flow, $\alpha=1$ whereas for a linear profile 
$\alpha=4/3$.  For sake of simplicity we will used in the following 
$\alpha=1$, the results being not sensitive to the choice of 
$\alpha$ as we will discuss in section 5.  The first force on the 
right-hand side is the gravity parallel to the plane which is the 
driving force.  The second term is the shear stress at the base: it is 
opposed to the motion, and is written as a friction coefficient $\mu$ 
times the vertical normal stress $\rho g h \cos \theta$.  The friction 
coefficient {\it a priori} depends on the local thickness $h$ and the 
local velocity $\|{\bf u}\|$.  The last term in eq.  \ref{mom} 
represents the pressure force linked to the thickness gradient.  The 
coefficient $K$ represents the ratio of the vertical normal stress to 
the horizontal one (\cite{savage89}).  For quasi-static deformation 
this coefficient can be calculated from standard Mohr-Coulomb 
plasticity model as done by Savage and Hutter (1989).  However, for 
dense granular flows on rough surfaces the material behaves more like 
a fluid.  Numerical simulations tend to show that the vertical and 
horizontal normal stresses are equal (\cite{prochnow00,ertas01}).  We 
then choose in the rest of the paper $K=1$.  This choice will be discussed 
in Section 5.

In order to apply those equations to describe granular flows, one has 
to give the expression for the friction law $\mu \left( h, \|{\bf u}\| 
\right)$. No theory exists for determining the bottom shear stress in the 
dense regime. A first approximation is to consider the friction 
coefficient as constant. This approximation works well 
when the surface is smooth but does not predict correct results 
for flows on surface whose roughness is of the order of the particle 
diameter. Here we use a more complex friction law based on 
previous experimental results obtained for steady uniform flows.

\subsection{The empirical friction law}

First, information about the friction mobilized in thin granular 
layer is given by the onset of the flow.  Experimental measurements have 
shown the existence of two critical angles: an initially static granular 
layer starts to flow when the inclination reaches a critical value 
$\theta_{start}$.  To stop the flow, one has to decrease the 
inclination down to a lower angle $\theta_{stop}$.  The important 
result for the case of a layer on rough inclines is that the two angles 
are function of the initial thickness $h$ of the layer as shown in 
Fig.3  
for the case of the glass beads used in our experiments.  The two 
curves $\theta_{stop}(h)$ and $\theta_{start}(h)$ have the same shape 
but are translated one from another by roughly one degree. The same 
behavior is observed with different materials on different 
roughness conditions (\cite{pouliquen96,daerr99}). In the following 
we defined the tangent of these angles 
$\mu_{start}(h)=\tan \left(\theta_{start}(h) \right)$ and 
$\mu_{stop}(h)=\tan \left(\theta_{stop}(h) \right)$.  We show below that the knowledge 
of these two functions is sufficient to define the empirical 
friction law $\mu \left( h, \|{\bf u}\| \right)$ in the whole range of 
velocity and thickness.

  \begin{figure}
  \centering 
  \includegraphics[scale=0.6]{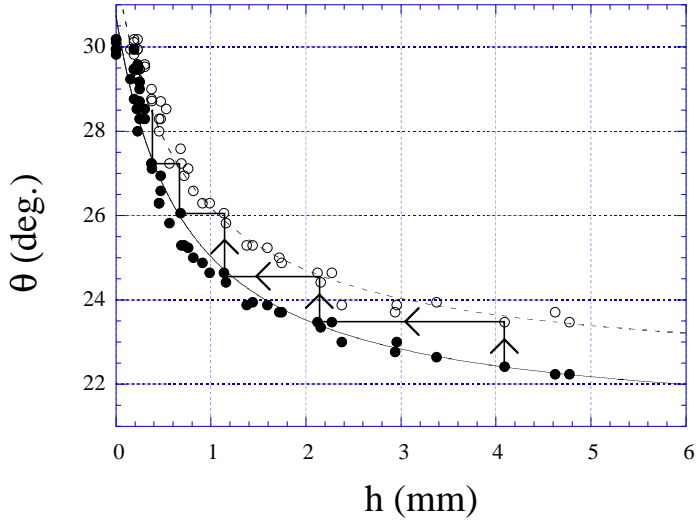}
  \caption{Variation of the starting ($\circ$)  and stopping ($\bullet$) angles 
  as a function of the thickness $h$. The arrows indicate the way they 
  are measured: starting with a uniform layer $h$, $\theta$ is 
  increased 
  up to the point where an avalanche occurs. After the avalanche, $h$ has 
  a lower value, $\theta$ is increased again and so on. In the $(h,\theta)$ plane 
  one oscillates between the starting and stopping curves.}
  \end{figure}

We have shown in a previous paper (\cite{pouliquen99a}) that 
the friction coefficient $\mu \left( h, \|{\bf u}\| \right)$ is 
related to the function $\mu_{stop}(h)$ by the relation
\begin{equation}
	\mu \left( h, \|{\bf u}\| 
\right)= \mu_{stop} \left(h \frac{\beta  \sqrt{g h}}{\|{\bf u}\|}\right)
\label{eq:mu1}
\end{equation}
where $\beta$ 
is a constant equal to 0.136 for glass beads, independent of the 
roughness conditions.

This non trivial result comes from properties observed for steady uniform flows
(\cite{pouliquen99a}). Steady uniform flows are controlled by the 
balance between friction and gravity which according to eq.\ref{mom} 
gives $\mu(h,\|{\bf u}\| )=\tan\theta$. To find the function $\mu$ one 
simply has 
to know how the mean velocity $\|{\bf u}\| $ varies with 
the two control parameters $h$ and $\theta$. Taking the inverse of the relation $\|{\bf u}\| (h,\theta)$ 
gives $\mu(h,\|{\bf u}\| )$. Our measurements have 
shown that 
 the mean velocity $\|{\bf u}\| $, the thickness of the granular layer $h$, and the 
inclination angle $\theta$ are related through the following relation:
\begin{equation}
	¥\frac{\|{\bf u}\| }{\sqrt{gh}}=\beta \frac{h}{h_{stop}(\theta)}.
	\label{eq:hstop}
\end{equation}
 The function $h_{stop}(\theta)$ is the 
thickness of the deposit left by a steady uniform flow at 
the inclination $\theta$. It is simply the inverse function of 
$\theta_{stop}(h)$. The expression of the 
friction coefficient eq. \ref{eq:mu1} is then obtained by substituting
 $\theta$ by $\tan^{-1}(\mu)$ in the velocity expression
eq.\ref{eq:hstop}.

However, the scaling law \ref{eq:hstop} is only valid for a Froude 
number defined as $Fr=\|{\bf u}\|  /\sqrt{gh}$ greater than $\beta$: no steady 
uniform flow is observed with a lower Froude number, i.e. with a 
thickness $h$ less than $h_{stop}$. It follows 
that the friction law  \ref{eq:mu1} is only valid for $Fr>\beta$.

For $Fr<\beta$ we have no information about the friction law.  
Experiments on friction between solids (\cite{heslot94}) or between a rough plate and a granular 
layer (\cite{nasuno97}) in the low velocity regime have shown a complex 
and rich dynamics not yet fully understood. As a first approximation, 
the important point is that the friction coefficient first decreases 
when increasing the velocity down to a minimum where it starts to increase. 
The velocity declining part is a source of instabilities 
(\cite{heslot94,baumberger94}). The fact that no steady 
uniform flow is observed for $Fr<\beta$ in our granular system suggests that the friction 
coefficient in this range decreases when increasing the Froude number. For  $Fr<\beta$ we then 
assume that the friction coefficient $\mu \left( h, \|{\bf u}\| 
\right)$ tends to the static friction coefficient $\mu_{start}(h)$ 
when velocity goes to zero.  In between $Fr=0$ and $Fr=\beta$, we 
simply
extrapolate by a power function characterized by a power $\gamma$. 

The 
final expression for the friction coefficient can then be written as 
follows (in terms of the thickness $h$ and the local Froude number 
$Fr=\|{\bf u}\|/\sqrt{gh}$ instead of $h$ and $\|{\bf u}\|$ ):

\vspace{0.3cm}
If $Fr>\beta$:
\begin{equation}
 \mu(h,Fr)= \mu_{stop}\left(h \frac{\beta }{Fr} \right)
\label{mu1}
\end{equation}
 \hspace{2cm} 

if $0<Fr<\beta$:
\begin{equation}
	 \mu(h,Fr)=\left(\displaystyle\frac{Fr}{\beta}\right)^{\gamma}
	 \left( \mu_{stopstart}(h)-\mu_{start}(h) \right)+\mu_{start}(h)
	 \label{mu2}
\end{equation}
\hspace{2cm} 
\vspace{0.5cm}

if $Fr=0$:
\begin{equation}
	\mu(h,0)=min\left(\mu_{start}(h),\|\tan \theta {\bf 
	e}_x - K {\bf \nabla} . h \| \right).
\end{equation}

\vspace{.3cm}

The last expression is just to ensure that, when the material is static, 
the friction exactly balances the other forces unless the total force 
reaches the threshold value given by the static friction coefficient. 

The crucial point is that the above friction force is {\it quantitatively} 
determined by the two functions $\mu_{stop}(h)$ and $\mu_{start}(h)$ 
that can be easily measured in the experiments. For the glass beads used in the 
experiments the best fits (Fig. 3) are given by:

\begin{equation}
		\mu_{stop}(h)=\tan\delta_{1}+\left(\tan\delta_{2}-\tan\delta_{1}\right) 
	\displaystyle\frac{1}{h/L +1}
\end{equation}

\begin{equation}
	\mu_{start}(h)=\tan\delta_{3}+\left(\tan\delta_{2}-\tan\delta_{1}\right) 
	\displaystyle\frac{1}{h/L +1}
\end{equation}

with $\delta_{1}=21^{o}$, $\delta_{2}=30.7^{o}$, $\delta_{3}=22.2^{o}$, 
and $L=0.65$mm.

The only unknown coefficient in the friction law is the power $\gamma$ 
of the extrapolation at low Froude numbers.  We have checked that the 
prediction of the model are not sensitive to its value as long as 
$\gamma$ is less than $10^{-2}$.  In the simulation $\gamma$ is chosen equal 
to $10^{-3}$. The sensitivity to the value $\gamma$ will be discussed 
in Section 5. The friction coefficient $\mu(h,Fr)$ given by 
equation 3.5-3.9 is plotted in Fig. 4 as a 
function of the Froude number for 
different thicknesses. The reader has to keep in mind that only the 
part in between $Fr=0$  and $Fr=\beta$ is arbitrarily chosen. The value 
at $Fr=0$ (stars) and for $Fr>\beta$ is quantitatively given by the 
measurement of the function  $\mu_{start}(h)$ and $\mu_{stop}(h)$. 
The discontinuity in the slope for $Fr=\beta$ is not 
physical and comes from the simple extrapolation we have chosen for 
the friction coefficient at low Froude number. We have checked that a smoother 
extrapolation around $Fr=\beta$ gives the same predictions.

The choice of the friction coefficient described above is such that it 
predicts the correct velocity for steady uniform flows, and the 
correct hysteresis for uniform layers (starting and stopping angles).  The 
question is whereas this friction law is sufficient to predict {\it 
quantitatively} more complex avalanching situations when it is introduced in the 
depth averaged equations \ref{mom}. 

  \begin{figure}
  \centering 
  \includegraphics[scale=0.5]{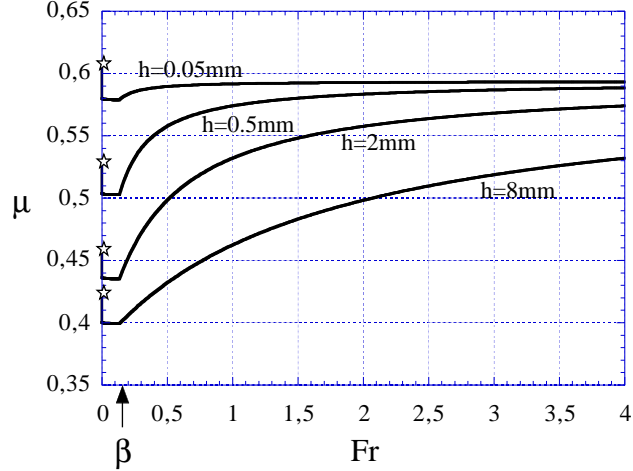}
  \caption{Friction coefficient $\mu(h,Fr)$ as a function of the Froude number 
  $\|u\|/sqrt{g h}$ for 
  different thickness $h$. The stars gives the values for $Fr=0$.}
  \end{figure}

\subsection{Numerical methods}

In order to numerically solve the depth averaged equations we have 
used two methods depending on the configuration of interest.  For the 
spreading of the mass on the empty plane, a lagrangian method has 
been chosen as it allows to easily treat parts where the thickness is zero. 
The method consists in discretizing the medium in elements which move 
and deform along with the flow.  
For the release of the mass on an initially static granular layer 
where shocks are likely to develop, we 
have written a shock capturing code based on a first order Godunov 
scheme. Both numerical schemes are described in detail in the 
appendix.

\section{Results}
\subsection{Release of a mass on a rough surface.}
We first have studied the spreading of the mass on the rough surface 
free of particles.  
We have systematically measured the time evolution of the shape of 
the mass $h(x,y,t)$ for different inclinations and initial volumes and 
compared with the predicted evolution given by the simulation of the 
depth averaged equations. No fit parameter exists in the simulation as 
soon as the functions $\mu_{stop}(h)$ and $\mu_{start}(h)$ are determined. 

The time evolution of the spreading for the medium cap at 
$\theta=23^{o}$ is presented in Fig. 5. The first row of figures shows 
contours of constant thickness measured in the experiments at five 
different times.  When the material starts to flow the mass rapidly 
spreads in both directions before it slowly stretches in the slope 
direction.  The rear part of the mass stops (around t=2s), whereas the front 
forms a bump which propagates down the slope.  The mass ultimately stops in a 
tear like shape.  This scenario is well predicted by the theory as 
shown by the second row of figures in Fig.5 giving the numerical predictions 
at the corresponding times.  The last series of figures is a comparison of the 
thickness profiles along the longitudinal symmetry axis $y=0$. Satisfactory 
agreement is obtained between theory and experiments from the 
initiation of the flow, up to the deposit. The presence of a thicker 
part flowing at the front when the rear part has already stops 
(t=2.4s) is well 
reproduced by the theory.

  \begin{figure}
  \centering 
  \includegraphics[scale=0.4]{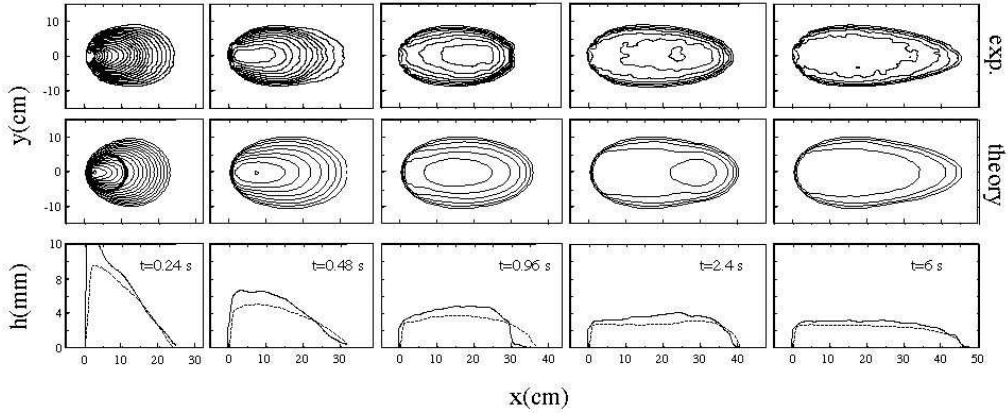}
  \caption{Time evolution of the granular mass for the medium cap $\theta=23^{o}$. First 
  row: experiments. Second row: simulations; Contour of constant thickness 
  every 0.5mm.  Third row: thickness profiles along $y=0$; solid line: 
  experiments; dotted line: simulation. The bold circle on the first plot of 
  the simulation represents the initial position of the cap.}
  \end{figure}

The agreement remains correct for different inclinations as shown in 
Fig. 6 showing the deposits measured in the experiments and predicted 
by the simulations for different 
inclination angles. Although the shape of the deposit looks similar 
from one inclination to another, we have not been able to put in 
evidence any similarity shape solutions.

  \begin{figure}
  \centering 
  \includegraphics[scale=0.55]{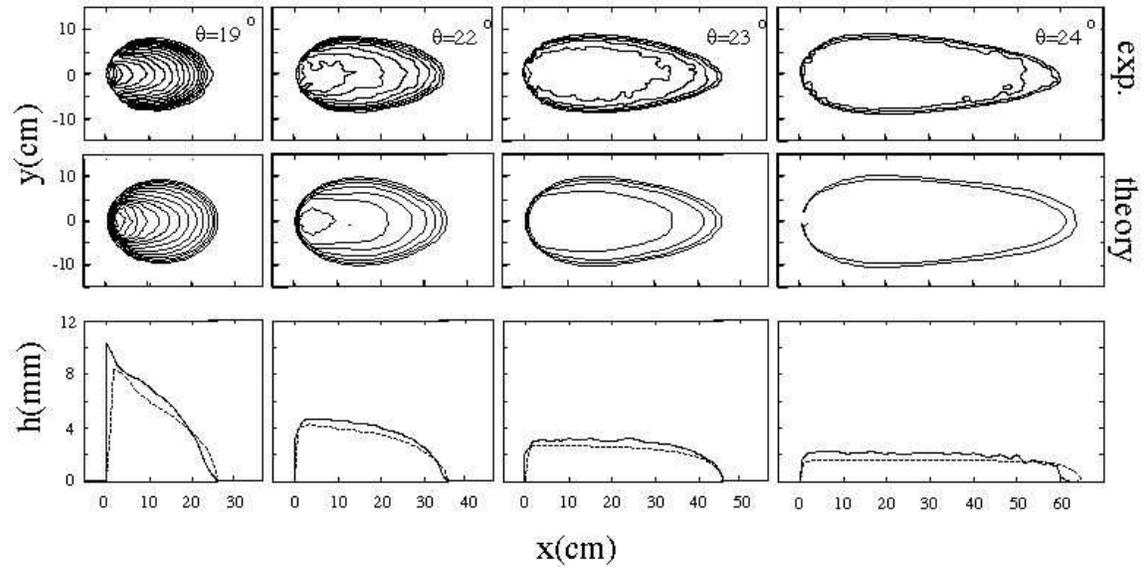}
  \caption{Comparison between experiments (first row) and simulations 
  (second row) for the deposit obtained at different inclinations with 
  the medium cap.  Contours of constant thickness 
  every 0.5mm.}
  \end{figure}

The effect of the initial mass has also been studied and the results are 
presented in Fig. 7. We have plotted the maximum width $W$ and 
length $L$ of the 
deposit for the three caps we used, as a function of the inclination. 
The lines correspond to the prediction of the model, the markers to 
the experimental observations. The agreement is good for the 
runout $L$ of the deposit, whereas the lateral spreading $W$ is overestimated 
by approximately 20\%. 

It is interesting to compare the above results with the prediction of the 
simple model using a constant friction coefficient (\cite{savage89}). In this case, no 
deposit is predicted as soon as the inclination of the plane is higher than the 
friction angle. The whole material flows down the plane. For inclination 
lower than the friction angle, the pile spreads and stops when 
its free surface makes an angle with the horizontal equal to the 
friction angle. In any case, it is not possible with a constant friction 
law to predict a deposit whose free surface is parallel to the plane as 
observed in the experiments.   

  \begin{figure}
  \centering 
  \includegraphics[scale=0.5]{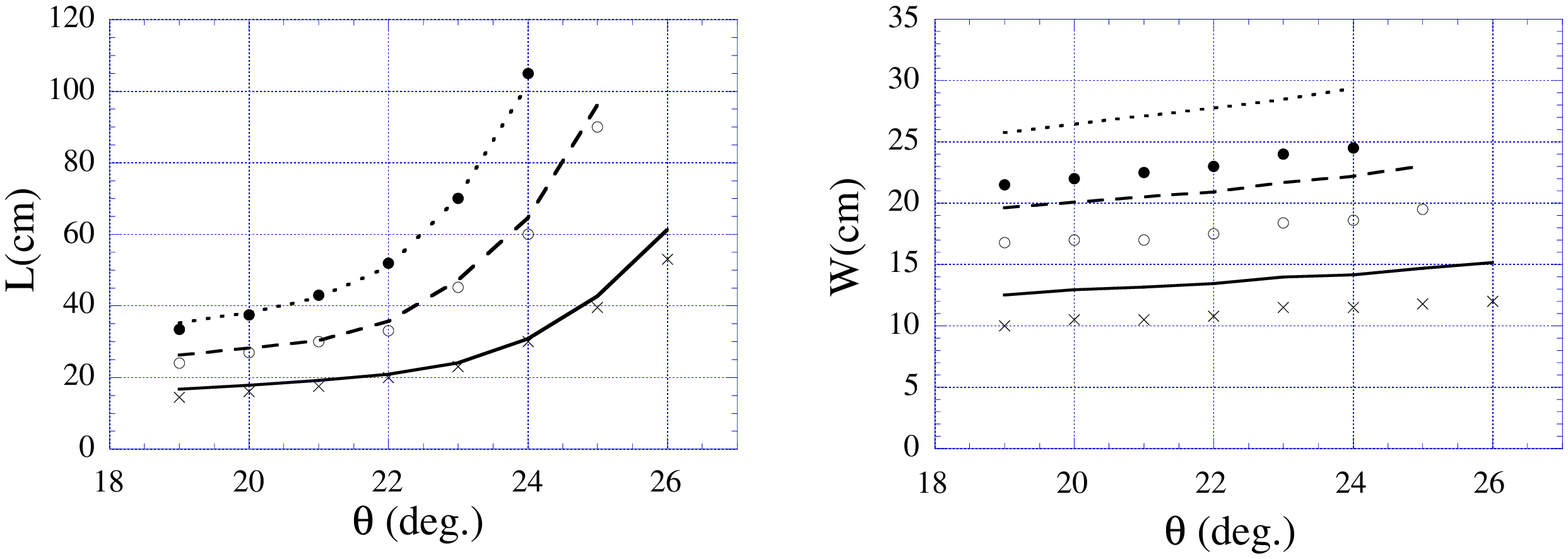}
  \caption{Comparison between experiments (markers) and simulation 
  (lines) for the maximum length $L$ and width $W$ of the deposit as a 
  function of $\theta$ and for three different caps; dotted 
  lines and black circles: large cap; broken lines and white circles: 
  medium cap; Solid lines and crosses: small cap.}
  \end{figure}

\subsection{Release of a mass on an initially static layer of material.}

The second set of experiments consists in releasing the mass on a 
 static layer of granular material. We first prepare the 
layer by creating a steady uniform flow at an angle $\theta$ and by
suddenly stopping the supply. A uniform deposit of thickness 
$h_{init}=h_{stop}(\theta)$ lies on the bed. We then carefully put the 
spherical cap on the layer and filled it with the beads. The mass is then 
released at t=0s. Result are presented in Fig. 8 for $\theta=23^{o}$ and 
for the small cap. The first row corresponds 
to the experimental measurements of the free surface 
represented by contours of constant thickness. The second row are the 
corresponding predictions of the simulation. The 
third row of figures gives the thickness profiles along the 
longitudinal symmetry axis y=0. 

The mass when released rapidly spreads in both directions up to a 
point when it reaches a drop shape which propagates down the slope 
without significant deformations.  This drop is a wave which put into 
motion static material at the front and leave static material at the 
rear.  The front is very sharp and the thickness profile along $y=0$ in 
the stationary regime is triangular as observed previously 
 by Daerr (2001).  This wave propagates faster than 
the material front in the previous deposit experiments described in section 4.1.

  \begin{figure}
  \centering 
  \includegraphics[scale=0.52]{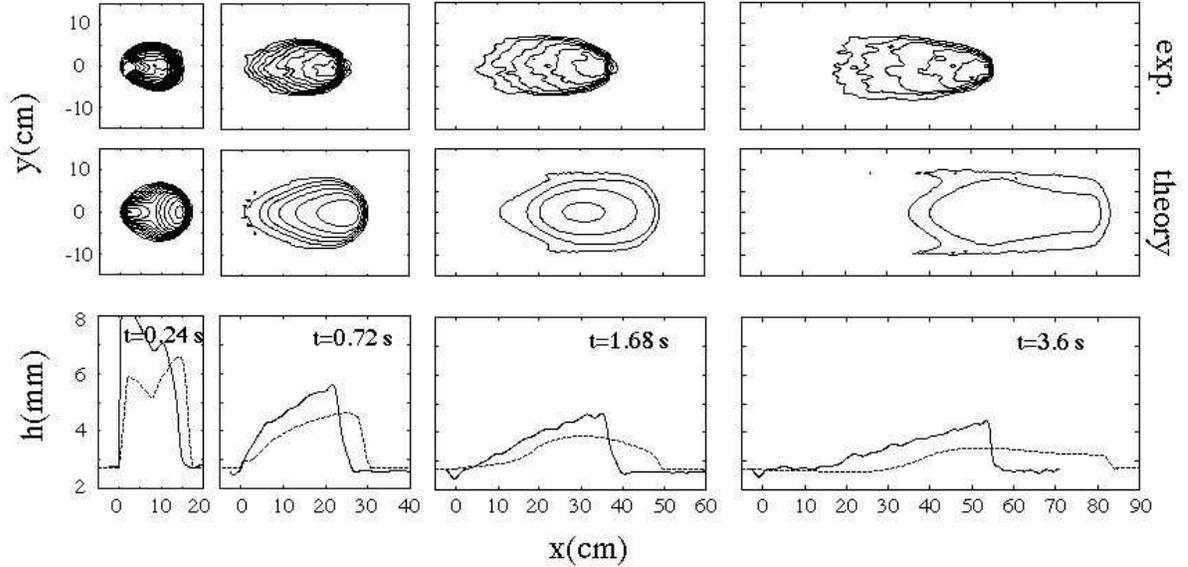}
  \caption{Time evolution of the granular mass released from the small cap 
  on an initial layer of thickness $h_{init}=2.7mm$, 
  $\theta=23^{o}$. First 
  row: experiments. Second row: simulations; Contour of constant thickness 
  every 0.3mm (the external contour line is h=3mm). Third row: thickness profiles 
  along $y=0$; solid line: 
  experiments; dotted line: simulations.}
  \end{figure}

The simulation qualitatively predicts the same dynamics.  The mass 
first spreads in both direction before reaching a roughly constant shape after 
about 4s (not shown in Fig. 8)  which propagates down the slope as a 
wave.  In front of the wave and behind it the material is at rest.  
Whereas the rapid initial spreading (t<0.72s) seems to be 
quantitatively correctly predicted by the model, the long time 
evolution is no longer in agreement with the experiments.  The model 
predicts a saturated thickness at the front thinner than the one 
observed in the experiment and the predicted propagation velocity is 
higher than the one observed. The front of the wave observed in the 
experiment is also more shocked than the one predicted by the 
simulations.  We have carried out experiments at 
different inclinations and with the other caps, which reveal the same 
discrepancy: the depth averaged model predicts always a propagating wave too 
rapid and too thin compared to the one observed in the experiments.

\section{Sensitivity of the model to the parameters}

From the two configurations we have investigated we can conclude that the 
depth averaged model quantitatively predicts the formation of deposit 
on the plane when the surface is free of particles but fails in 
predicting the correct avalanche propagation when a static layer is present. 
In order to better understand the limits of the depth averaged 
approach we have studied the sensitivity of the simulation to the 
different parameters introduced in the model. We will then discuss the 
role of the parameters we have not measured: $\alpha$ in the 
acceleration term, $K$ in the pressure term, $\gamma$ the power of the 
extrapolation at low Froude number.

\subsection{Role of $\alpha$}

The parameter $\alpha$ in the acceleration term in Eq. \ref{mom} is 
obtained when integrating the 3D conservation equations to derive the 
depth averaged equation.  It depends on the assumed velocity profile 
across the layer.  In the simulation presented above $\alpha$ was 
taken equal to 1 corresponding to a plug flow.  For a linear profile 
$\alpha$ should be taken equal to 4/3.  For granular flow on rough 
inclined plane, previous results suggest that the profile is closer to 
a linear one than to a plug flow (\cite{prochnow00,ertas01}).  However, the results of the 
simulation are not very sensitive to the value of $\alpha$.  For 
example, simulations with $\alpha=4/3$ and $\alpha=1$ for the case of 
Fig. 8 gives a spatio-temporal evolution which differs only by a factor 
2\% for the thickness and velocities.  The reason why the model is so 
insensitive can be understood by comparing the relative magnitude of 
the forces acting on the material.  Except at the first instants of 
the spreading, the acceleration is negligible and the friction force 
balances the gravity forces. In other term, inertia is negligible in 
the flow regime investigated in our experiments.  This can be 
quantitatively estimated by locally measuring the ratio between the acceleration and the friction 
force defined by the dimensionless number $R_{f}$:
\begin{equation}
	R_{f}(x,y,t)=\frac{\| \tan \theta {\bf e}_x - \mu \left( h, \|{\bf u}\| \right) 
\displaystyle\frac{{\bf u}} { \|{\bf u}\|} - K {\bf 
\nabla}.h \|}{\mu \left( h, \|{\bf u}\| \right)}.
\end{equation}.

In the case of the deposit experiment of Fig. 5, $R_{f}$ is maximum during 
the initial rapid spreading where it reaches the value 0.4 (t<0.5s) and 
rapidly decreases up to 0.02 everywhere for t>0.8s.  In the case of 
the flow on a static layer (Fig. 8), $R_{f}$ is less than 0.02 everywhere 
during the avalanche except at the front where material is put 
into motion. In this region $R_{f}$ reaches a higher value around 0.1 .  We can then 
conclude that inertial effects are negligible for the dense granular 
flow we are studying.  The choice of $\alpha$ has then no significant 
influences on the prediction of the depth averaged model for the 
configuration of interest in this study. 

\subsection{Role of the pressure coefficient $K$}

A second parameter of the model is the pressure coefficient $K$ in 
front of the thickness gradient term of Eq. \ref{mom}.  This 
coefficient represents the ratio of the horizontal normal stress to the 
vertical one.  In the results presented above we have used $K=1$ 
corresponding to an isotropic pressure.  This choice has been 
motivated by several numerical results showing a small difference 
between the two normal stresses (\cite{prochnow00,ertas01}).  
Moreover, we observe that $K=1$ gives the best agreement for the 
predicted deposits in section 4.1.  An example of the prediction with 
another value of $K$ is shown in Fig 9.  We have used the value 
$K=\frac{1+\sin^{2}\delta_{1}}{1-\sin^{2}\delta_{1}}$ corresponding to 
the Mohr Coulomb prediction when the basal friction is equal to the 
internal friction coefficient taken equal to $\tan\delta_{1}$ 
(\cite{savage89}).  For 
our glass beads we obtained $K=1.29$.  Using this value, the predicted 
deposit is wider and shorter than the one experimentally observed.  
For the case of the flow on the static layer of Fig. 8 the higher 
value of $K$ predicts also a larger spreading and does not influence 
the front propagation.  Changing $K$ does not provide better agreement 
for the propagation of the wave.

  \begin{figure}
  \centering 
  \includegraphics[scale=0.4]{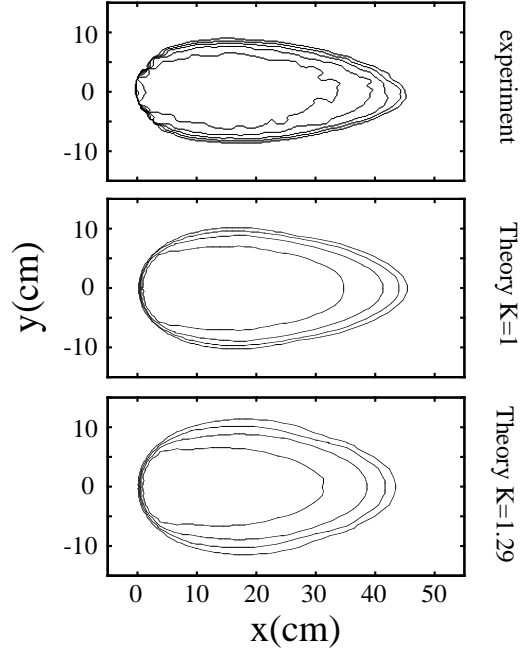}
  \caption{Deposits of the granular mass released from the medium cap 
  $\theta=23^{o}$. a) Experiment; b) Simulation with $K=1$; c) Simulation 
  with $K=1.29$, the value given by the Mohr-Coulomb plasticity criterium. 
  }
\end{figure}

\subsection{Role of the power extrapolation $\gamma$.}

The expression of the friction force in our model is quantitatively 
determined by the two functions $\mu_{stop}(h)$ and $\mu_{start}(h)$ 
except for the extrapolation at low Froude number.  The extrapolation 
we have chosen is characterized by a power function with an exponent 
$\gamma$.  In the simulation presented above, we have used 
$\gamma=10^{-3}$.  This low value insures a rapid increase of the 
friction coefficient close to $Fr=0$ (Fig.  4).  No influence of 
$\gamma$ is observed up to $10^{-2}$.  If one chooses a larger power 
($\gamma>10^{-2}$), the depth averaged model slightly underestimates 
the runout distance and overestimates the thickness of the deposit, especially 
at the rear where a small bump is predicted which is not observed.  
However, the effects are small. For example, for $\theta=23^{o}$ the 
case $\gamma=1$ predicts a runout distance $L=41cm$ and a maximum deposit
thickness of $h=3.3mm$ whereas with $\gamma=10^{-3}$ one get $L=45cm$ 
and $h=2.8mm$.

For the case of the flow on a static layer, changing the value of $\gamma$ 
does not significantly changes the front velocity in the Fig. 
8.  

In conclusion it seems that the discrepancy observed in Fig. 8 between the model and the 
measurement for the case of the kinematic wave moving on a static 
layer can not be erased by tuning some of the parameters of 
the model. 

\section{Discussion and conclusion}

We have shown in this paper that the flow of granular material along rough 
surfaces can be correctly described by introducing a relevant 
friction law in a depth averaged description.  
The spreading of a mass on a rough slope free of particles is 
{\it quantitatively} predicted by the model from the release up to the deposit.  
The interesting point is that the friction law introduced in the model 
for describing the interaction between the flowing granular layer and 
the rough surface  is mainly quantitatively 
determined by the dependence of the stopping and 
starting angle with the thickness of the layer. Measuring these angles 
is enough to predict the velocity, the shape and the run-out of the 
mass. 

However, we have shown in this study that the agreement is no longer correct when 
studying the release of the mass on a uniform granular layer initially 
static on the surface.  The formation of a wave which puts into motion static 
material at the front and leaves material at the rear is predicted by 
the depth averaged model, but the corresponding wave velocity and amplitude 
are not quantitatively in agreement with the measurements. 

Several points could explain the discrepancy. First, the friction law 
we have used is too simple in the low velocity 
regime and could be responsible for the wrong 
description of the front. However, we have not found a description 
which could capture the correct avalanche propagation. Secondly, at 
the front 
between the flowing part and the static part, the long wave assumption 
breaks down. The thickness of the layer varies on length scale of the 
order of the thickness. Thirdly, an underlying assumption in the 
depth averaged equations is that the time variation of the properties 
at one point are slow enough that the velocity profile across the layer 
 can be considered as fully developed. This assumption 
 certainly breaks down for an avalanching front putting into motion 
 some static material. The whole layer does not 
 instantaneously start to flow but material at the free 
 surface first moves, putting into motion the deeper layers. For such 
 configurations, more complex approaches taking into account two 
 layers, one for static grains, the other one for flowing grains 
 could be more relevant (\cite{bouchaud94,boutreux98,douady99}). The 
 question of the relevant friction law which should be used at the 
 interface between the two layers is an open question. The one 
 proposed in this study is perhaps not appropriate as it takes into 
 account the influence of a rigid wall.  
 
 Geophysical flows are more complex than the simple avalanches of 
 glass beads studied 
 in this work. One can legitimately wonder to what extent similar 
 approaches could be relevant for describing natural events. A first important 
 difference concerns the nature of the material. Geophysical events 
 involve complex heterogeneous granular material made of particles of 
 various sizes ranging from microns to meters often mixed with fluids. 
 In such polydispersed material it is well known that segregation 
 occurs with the large particles rising up to the free surface. The 
 influence of the segregation process on the spreading of the mass 
 and on the traveling distance of the avalanche
 are questions which represent works for future investigations. 
 
 Another difficulty for describing natural events is the complex 
 topography.  However, depth averaged equations can be derived for non 
 uniform slopes, by introducing a local inclination that varies from 
 one point to another. Works have been carried out in this direction 
 on real topographies (\cite{naaim97,heinrich01}). The results are 
 promising showing that the depth averaged approach is certainly relevant 
 when no mass exchange exists between the flowing mass and the 
 substrate. However, the choice of the friction law for geophysical material 
 remains an open issue.
 
 \acknowledgements
 This research was supported by the Ministre Franais de la 
 Recherche (ACI blanche $\# 2018$). Many thanks goes to R. Saurel for 
his help in developping the Godunov code. Discussions with J. 
Misguisch were helpful for developping the MoirŽ method. We thank F. Ratouchniak.
for his technical 
assistance.  
 
\appendix

\section{Lagrangian scheme}

The lagrangian method for simulating the depth averaged equations is 
very similar to the one described by Wieland et al (\cite{wieland99}). The granular mass 
is discretized into a finite number of parallepipedic elements which 
move and deform along with the flow. Velocities are defined at the node 
$i$ of the grid and thickness at the middle of the element. The time 
evolution is obtained as follows:

i) 
The position ${\bf x}_{i}^{t+1}$ of the node i at time t+1 is 
evaluated by:
$${\bf x}_{i}^{t+1}= {\bf x}_{i}^{t}+\Delta t {\bf u}_{i}^{t}$$
where ${\bf u}_{i}^{t}$ is the velocity. 

ii) The thickness at the center of the element is computed using the 
mass conservation:
$$ h_{i}^{t+1}= V_{i}/ S_{i}^{t+1},$$
where $V_{i}$ is the volume of element $i$ which remains constant during 
the flow, and $S_{i}^{t+1}$ is the surface of element $i$ given by 
the new positions of the four corners of the element. 

iii) The velocities are updated as follows:
$$ {\bf u}_{i}^{t+1}= {\bf u}_{i}^{t} + \Delta t {\bf F}_{i}^{t}$$
where ${\bf F}_{i}^{t}$ are the forces at the node corresponding to the right hand 
side of eqs \ref{mom}. To evaluate ${\bf F}_{i}^{t}$ one needs to know the 
thickness and thickness gradient at the nodes. They are evaluated 
using standard finite element interpolation (\cite{grandin86}). Note 
that with this lagrangian formulation the coefficient $\alpha$ in the 
acceleration term in eq. \ref{mom} is equal to $1$. 

The $35\times 35$ elements are initially disposed 
as in Fig. 10a and the thickness is initialized to fit the spherical cap 
used in the experiments. Initial velocities are zero and the time increment 
is set to $5.10^{-4}s$.

\section{Shock capturing scheme}

We have developed a second code to model the flow on an initially static 
material. It is based on a first 
order Godunov scheme 
(\cite{godunov59}) in order to be able to correctly capture shock propagation. The 
equations \ref{masscons} and \ref{mom} are rewritten in a conservative form:
$$
\displaystyle\frac{\partial  U}{\partial t} + 
\displaystyle\frac{\partial G^{x}}{\partial x}+
\displaystyle\frac{\partial  G^{y}}{\partial y} = F
$$

where $U$, $ G^{x}$, $ G^{y}$ and  $F$ are vectors
depending on the thickness $h$ and longitudinal and transverse 
velocities $u$ and $v$ as follows:

\vspace{.5cm}
$U=\left(
\begin{array}{c}
	h\\
	hu\\
	hv
\end{array}
\right)
$ , 
$G^{x}=\left(
\begin{array}{c}
	hu\\
	\alpha hu^{2}+Kh^{2}\\
	\alpha h u v
\end{array}
\right)
$ , 
$G^{y}=\left(
\begin{array}{c}
	hv\\
	\alpha h u v\\
		\alpha hv^{2}+Kh^{2}
\end{array}
\right)
$ , 

\vspace{.5cm}
$F=\left(
\begin{array}{c}
	0\\
	\rho g h \cos \theta \left( \tan \theta - \mu( h, Fr) 
\frac{u} { \sqrt{u^{2}+v^{2}}} - 
	K\frac{\partial h}{\partial x}\right) \\
	\rho g h \cos \theta \left( - \mu( h, Fr) 
	\frac{v} { \sqrt{u^{2}+v^{2}}} - 
	K\frac{\partial h}{\partial y}\right)
\end{array}
\right)
$
\vspace{.5cm}

The data are discretized on a regular grid. 
Knowing the vector $U^{t}_{i,j}$ at time $t$ and at the point $(i,j)$ one 
computes its value at time $t+\Delta t$ as follows:
$$
U^{t+\Delta t}_{i,j}=U^{t}_{i,j}-\frac{\Delta t}{\Delta 
x}\left(G^{x}_{i+1/2,j}-G^{x}_{i-1/2,j}\right) - \frac{\Delta t}{\Delta 
y}\left(G^{y}_{i,j+1/2}-G^{y}_{i,j-1/2}\right) + F_{i,j} \Delta t 
$$

where $G^{x}_{i+1/2,j}$ and $G^{y}_{i,j+1/2}$ are the horizontal and vertical fluxes calculated 
at the frontier between cells (Fig.10b). We use the HLL (\cite{harten83}) approximation to 
compute the fluxes at the frontier from the fluxes at the center of 
the cell:
$$
G^{x}_{i+1/2,j}=\displaystyle\frac{{C^{x,l}}_{i+1/2,j}G^{x}_{i,j}-{C^{x,r}}_{i+1/2,j}G^{x}_{i+1,j} 
+ 
{C^{x,l}}_{i+1/2,j}{C^{x,r}}_{i+1/2,j}(U^{t}_{i+1,j}-U^{t}_{i,j})}{{C^{x,l}}_{i+1/2,j}-{C^{x,r}}_{i+1/2,j}} 
,
$$
$$
G^{v}_{i,j+1/2}=\displaystyle\frac{{C^{y,l}}_{i,j+1/2}G^{y}_{i,j}-{C^{y,r}}_{i,j+1/2}G^{y}_{i,j+1} + 
{C^{y,l}}_{i,j+1/2}{C^{y,r}}_{i,j+1/2}(U^{t}_{i,j+1}-U^{t}_{i,j})}{{C^{y,l}}_{i,j+1/2}-{C^{y,r}}_{i,j+1/2}}.
$$
The wave velocities are estimated by the following choice proposed by 
Davies (1988):
$$
\begin{array}{l}
{C^{x,l}}_{i+1/2,j}=min(0,\alpha u_{i+1,j}-c^{x}_{i+1,j},\alpha u_{i,j}-c^{x}_{i,j}),\\
{C^{x,r}}_{i+1/2,j}=max(0,\alpha u_{i+1,j}+c^{x}_{i+1,j},\alpha u_{i,j}+c^{x}_{i,j}),\\
{C^{y,l}}_{i,j+1/2}=min(0,\alpha u_{i,j+1}-c^{y}_{i,j+1},\alpha u_{i,j}-c^{y}_{i,j}),\\
{C^{y,r}}_{i,j+1/2}=max(0,\alpha u_{i.j+1}+c^{y}_{i,j+1},\alpha u_{i,j}+c^{y}_{i,j}).\\
\end{array}
$$
with $c^{x}_{i,j}=\sqrt{(\alpha^{2}-\alpha)u^{2}_{i,j}+2Kh_{i,j}}$ and 
$c^{y}_{i,j}=\sqrt{(\alpha^{2}-\alpha)v^{2}_{i,j}+2Kh_{i,j}}$.

The simulations presented in the paper have been obtained with $\Delta 
x= \Delta 
y=0.15$ cm and  $\Delta 
t=5.10^{-4}$s. 

  \begin{figure}
  \centering 
  \includegraphics[scale=0.5]{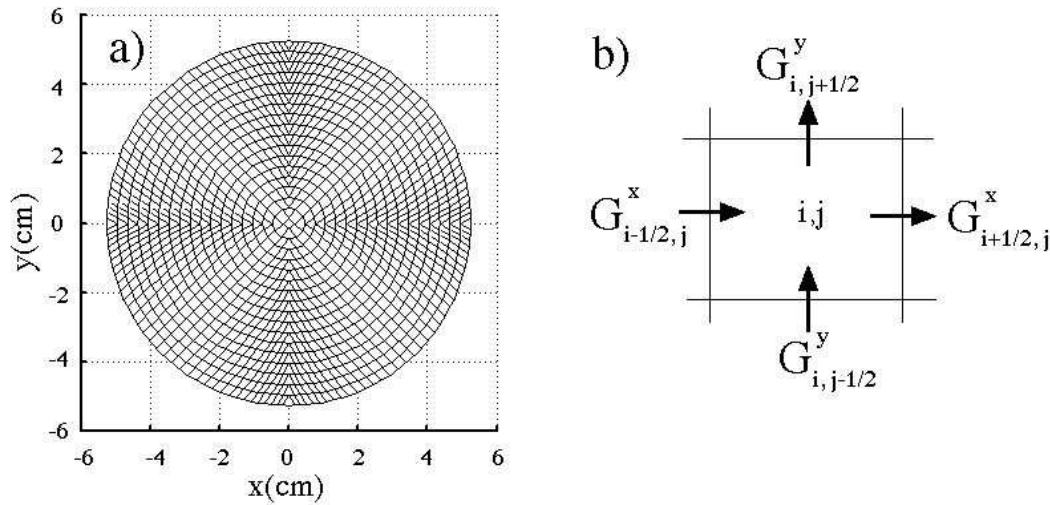}
  \caption{a) Initial position of the elements for the lagrangian 
  simulation for the medium cap; b) Sketch of one cell and the 
  corresponding  horizontal 
  and vertical fluxes for 
  the Godunov scheme.  
  }
\end{figure}

\newpage

\end{document}